\documentclass[aps,pra,twocolumn,superscriptaddress]{revtex4-1}

\usepackage{amsmath,amssymb,amsfonts}
\usepackage{graphicx}
\usepackage{siunitx}
\usepackage{natbib}
\usepackage{braket}
\usepackage{array}
\usepackage{multirow}

\usepackage{url}
\usepackage[breaklinks]{hyperref}
\hypersetup{
    unicode=true,
    colorlinks=true,
    linkcolor=blue,
    citecolor=blue,
    urlcolor=blue
  }
	\usepackage{breakurl}

\graphicspath{ {Figures/} }

\begin{document}

\title{Production of ultracold Cs$^*$Yb molecules by photoassociation}
\author{Alexander Guttridge}
\email{alexander.guttridge@durham.ac.uk}
 \affiliation{Joint Quantum Centre (JQC) Durham-Newcastle, Department of Physics, Durham University, South Road, Durham, DH1 3LE, United Kingdom.}
\author{Stephen A. Hopkins}
\affiliation{Joint Quantum Centre (JQC) Durham-Newcastle, Department of Physics, Durham University, South Road, Durham, DH1 3LE, United Kingdom.}
\author{Matthew D. Frye}
\affiliation{Joint Quantum Centre (JQC) Durham-Newcastle, Department of Chemistry, Durham University, South Road, Durham, DH1 3LE, United Kingdom.}
\author{John J. McFerran}
\affiliation{Department of Physics, University of Western Australia, 6009 Crawley, Australia}
\author{Jeremy M. Hutson}
\email{j.m.hutson@durham.ac.uk} \affiliation{Joint Quantum Centre (JQC)
Durham-Newcastle, Department of Chemistry, Durham University, South Road,
Durham, DH1 3LE, United Kingdom.}
\author{Simon L. Cornish}
\email{s.l.cornish@durham.ac.uk}
\affiliation{Joint Quantum Centre (JQC) Durham-Newcastle, Department of Physics, Durham University, South Road, Durham, DH1 3LE, United Kingdom.}

\begin{abstract}
We report the production of ultracold heteronuclear Cs$^*$Yb molecules through
one-photon photoassociation applied to an ultracold atomic mixture of Cs and Yb
confined in an optical dipole trap. We use trap-loss spectroscopy to detect
molecular states below the Cs($^{2}P_{1/2}$) + Yb($^{1}S_{0}$) asymptote. For
$^{133}$Cs$^{174}$Yb, we observe 13 rovibrational states with binding energies
up to $\sim$500\,GHz. For each rovibrational state we observe two resonances
associated with the Cs hyperfine structure and show that the hyperfine
splitting in the diatomic molecule decreases for more deeply bound states. In
addition, we produce ultracold fermionic $^{133}$Cs$^{173}$Yb and bosonic
$^{133}$Cs$^{172}$Yb and $^{133}$Cs$^{170}$Yb molecules. From mass scaling, we
determine the number of bound states supported by the 2(1/2) excited-state
potential to be 154 or 155.
\end{abstract}

\date{\today}

\maketitle

\section{Introduction}

Ultracold polar molecules are a promising platform for the study of new forms
of quantum matter \cite{Baranov2002,Buechler2007,Cooper2009}, cold controlled
chemistry \cite{Krems2008,Ospelkaus2010} and tests of fundamental physics
\cite{Flambaum2007,Meyer2009,Isaev2010,Hudson2011,Safronova2017}. The electric
dipole moment possessed by polar molecules can be exploited to engineer
controllable long-range dipole-dipole interactions, which have many
applications in quantum simulation \cite{Barnett2006,Gorshkov2011,Bohn2017},
quantum computation \cite{DeMille2002} and the study of quantum many-body
physics \cite{Carr2009,Baranov2012}. Many of these applications require gases
of ground-state molecules with high phase-space density confined in optical
traps or lattices. Whilst direct laser cooling of molecules has undergone
spectacular recent progress
\cite{Hummon2013,Steinecker2016,Kozyryev2017,Truppe2017}, the molecules
produced in these experiments are currently limited to low phase-space
densities. However, high-phase-space-density gases of ultracold molecules can
be produced from ultracold mixed-species gases of alkali-metal atoms using
magnetoassociation on a Feshbach resonance followed by optical transfer to
deeply-bound states.

High-phase-space-density gases of KRb \cite{Ni2008}, RbCs
\cite{Takekoshi2014,Molony2014}, NaK \cite{Park2015} and NaRb \cite{Guo2016}
molecules have been produced in the $^1\Sigma$ ground-state using this approach
and the first steps towards realising the richness of ultracold molecular
systems have been demonstrated using such bi-alkali molecules
\cite{Yan2013,Park2017,Blackmore2018}. At the same time, the quest for new species of
ultracold molecules possessing a magnetic dipole moment, in addition to an
electric dipole moment, has become a field of burgeoning interest, with both
$^2\Sigma$
\cite{Tassy2010,Hara2011,Pasquiou2013,Roy2017a,Flores2017,Witkowski2017} and
$^3\Sigma$ molecules \cite{Rvachov2017} being pursued. The additional degree of
freedom possessed by these molecules allows quantum simulation of a wide range
of two-dimensional lattice spin models \cite{Micheli2007} and tuning of
collisions and chemical reactions \cite{Abrahamsson2007}.

Following the success of the association technique in bi-alkali experiments,
the association of an alkali-metal atom and a closed-shell atom is a promising
approach for the production of $^{2}\Sigma$ molecules. Magnetoassociation of
such molecules is complicated in comparison to the bi-alkali case due to the
singlet ground state of the closed-shell atom which precludes the existence of
broad Feshbach resonances. However, the weak distance dependence of the
hyperfine coupling, caused by the proximity of the second atom, is predicted to
produce usable Feshbach resonances \cite{Zuchowski2010, Brue2012} in these
systems, with CsYb one of the most promising candidates \cite{Brue2013}. Such
resonances have recently been observed experimentally in the RbSr system
\cite{Barbe2017}, but magnetoassociation remains unexplored.

\begin{figure}
		\includegraphics[width=0.97\linewidth]{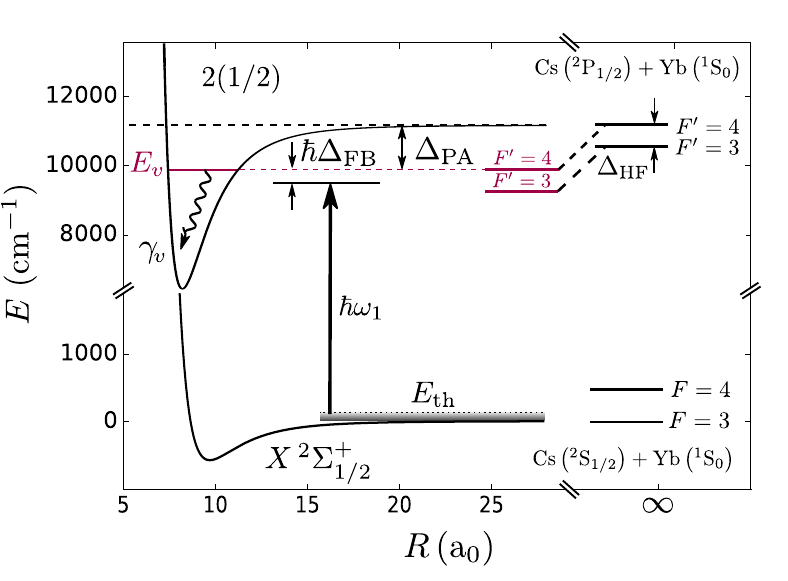}
	\caption{One-photon photoassociation. When $\hbar \omega_{1} = E_{v}$
($\hbar \Delta_{\mathrm{FB}} = 0$) a pair of colliding ground-state Cs and Yb
atoms are associated to form a CsYb molecule in a rovibrational level
of the electronically excited 2(1/2) molecular potential. The binding energy of
this rovibrational level, $\Delta_{\mathrm{PA}}$, is measured with respect to
the Cs $D_{1}$ line which the 2(1/2) potential approaches asymptotically. The
molecular curves plotted here are adapted from Ref.\ \cite{Meniailava2017}. The
hyperfine splitting shown on the right is not to scale. }
	\label{fig:PAdiagram}
\end{figure}

Light-assisted techniques such as photoassociation (PA) \cite{Jones2006} and
stimulated Raman adiabatic passage \mbox{(STIRAP)} \cite{Bergmann1998} offer
alternative approaches for the production of ground-state molecules in these
systems which are not reliant on the existence of suitable Feshbach resonances.
Photoassociation is a technique where a colliding atom pair is excited to a
rovibrational level of an excited molecular potential, forming an excited
molecule. The subsequent decay of the excited molecule is determined by the
Franck-Condon factor (FCF), which dictates the branching ratios for molecular
decay into energetically lower states, including the continuum. By choosing an
excited vibrational level with a favourable Franck-Condon overlap with the
ground state, photoassociation can be used as a method of producing
ground-state ultracold molecules
\cite{Kerman2004,Sage2005,Deiglmayr2008,Zabawa2011,Altaf2015}. Alternatively,
the coherent transfer of a colliding atom pair to a bound vibrational level of
the molecular ground state is also possible, as has been investigated in
Sr$_{2}$ \cite{Stellmer2012,Ciamei2017}. The two techniques can be combined
using photoassociation to populate a high-lying vibrational level followed by
coherent transfer to the absolute ground state \cite{Aikawa2010}. The first
step towards identifying viable routes for the creation of molecules using
these all-optical approaches involves sensitive photoassociation measurements
of near-threshold bound states to determine the long-range potential of the
excited molecular state. This technique is illustrated in Fig.\
\ref{fig:PAdiagram} for CsYb and is explored in this work.

In this paper we report the production of ultracold heteronuclear Cs$^*$Yb
molecules using one-photon PA applied (initially) to an ultracold atomic
mixture of $^{133}$Cs and $^{174}$Yb confined in an optical dipole trap (ODT).
We present measurements of the binding energies of rovibrational states up to
500\,GHz below the Cs($^{2}P_{1/2}$) + Yb($^{1}S_{0}$) asymptote. The
electronic state at this threshold is designated 2(1/2) to indicate that it is
the second (first excited) state with total electronic angular momentum
$\Omega=1/2$ about the internuclear axis. It correlates at short range with the
$1\,^2\Pi_{1/2}$ electronic state in Hund's case (a) notation
\cite{Meniailava2017}, but at long range the $^2\Pi_{1/2}$ and $^2\Sigma_{1/2}$
states are strongly mixed by spin-orbit coupling.

We fit an extended version of the Le\,Roy-Bernstein near-dissociation expansion
formula to the measurements and characterize the long-range potential in the
2(1/2) excited state. We investigate the role of hyperfine coupling in Cs$^*$Yb
molecules by studying the hyperfine splitting of the observed lines and show a
dependence on the internuclear separation. Finally, we expand the scope of our
investigation by measuring the PA spectra of an additional 3 CsYb isotopologs,
$^{133}$Cs$^{173}$Yb, $^{133}$Cs$^{172}$Yb and $^{133}$Cs$^{170}$Yb. Using mass
scaling, we determine the number of bound states supported by the 2(1/2)
molecular potential. These results represent a critical first step towards the
coherent production of molecules in the electronic ground-state by a two-photon
process.

\section{Experimental Setup}

Photoassociation measurements are typically performed in either a
magneto-optical trap (MOT) or an optical dipole trap (ODT). We use an ODT as
our experiment employs a single Zeeman slower that prevents continuous loading
of Cs and Yb into a dual-species MOT \cite{Kemp2016,Hopkins2016}. The use of an
ODT has the advantage that the internal states of the atoms are better defined,
the temperature is lower and the interspecies density is higher than in typical
MOT experiments. However, measurements in the ODT are performed using
destructive absorption imaging to determine the number of atoms remaining after
exposure to the PA light. PA spectra must therefore be built up by repeating
the experiment multiple times whilst stepping the frequency of the PA light.
This makes broad frequency scans much more time-consuming in comparison to MOT
measurements where the MOT fluorescence can be continuously monitored as the PA
laser frequency is scanned.

The ODT used in this work is formed from the output of a broadband fibre laser (IPG YLR-100-LP) with a wavelength of $1070(3) \,$nm, and consists of two beams crossed at an angle of $40 ^{\circ}$ with waists of $33(4) \, \mu $m and $72(4) \, \mu$m. The measured Yb (Cs) trap frequencies are 240 (750) Hz radially and 40 (120) Hz axially. The trap depths for the two species are $U_{\mathrm{Yb}} = 5 \, \mu$K and $U_{\mathrm{Cs}} = 85 \, \mu$K. We typically load the ODT with a mixture of \mbox{$8 \times 10^{5}$ $^{174}$Yb} atoms at $T_{\mathrm{Yb}} = 1 \, \mu$K in the $^{1}S_{0}$ ground state and \mbox{7 $\times 10^{4}$} Cs atoms at $T_{\mathrm{Cs}} = 6 \, \mu$K  in the absolute ground state $^{2}S_{1/2} \, \ket{F=3,m_{F}=+3}$. A detailed description of the experimental apparatus and the routine for the preparation of this mixture is given in Refs.\ \cite{Kemp2016,Hopkins2016,Guttridge2016,Guttridge2017}.

The PA light is derived from a Ti:Sapphire laser (\mbox{M Squared SolsTiS}),
the main output of which is passed through an acousto-optic modulator for fast
intensity control and coupled into a fibre which carries the light to the
experimental table. The PA light is focused onto the trapped atomic mixture
with a waist of $150 \, \mu$m and is polarized parallel to the applied magnetic
field in order to drive $\Delta m_F=0$ transitions. The hyperfine structure of
the weakly bound molecular states is similar to that of the atomic state (see
Fig.\ \ref{fig:PAdiagram}). The strengths of transitions to these molecular
states are dictated by the dipole matrix elements as in the atomic case
\footnote{In addition to dipole matrix elements, the strengths of molecular
transitions are subject to further effects such as the Franck-Condon overlap
and rotational couplings}. The choice of polarization allows the excitation to
molecular levels in both hyperfine manifolds.

The frequency of the PA light is both stabilized and calibrated using a
high-finesse optical cavity, the length of which is stabilised to a Cs atomic
transition using the Pound-Drever Hall method \cite{Drever1983}. PA light sent
to the cavity passes through a broadband fibre electro-optic modulator (EOM)
(EOSPACE PM-0S5-10-PFA-PFA-895) modulating the light with frequency sidebands.
We utilize the `electronic sideband' technique \cite{Thorpe2008,Gregory2015} to
allow continuous tunability of the PA laser frequency; by stabilising one of
the sidebands to a cavity transmission peak, the frequency of the carrier may
be tuned over the $748.852(5) \,$MHz free spectral range (FSR) of the cavity by
changing the modulation frequency applied to the EOM. Precise frequency
calibration with respect to the Cs $D_{1}$ transition is then achieved by
counting cavity fringes from the $D_{1}$ transition and including the rf
modulation offsets of the carrier. In practise a commercial wavemeter (Bristol
671A) is used to identify the specific cavity fringe used to stabilise the PA
laser frequency.

\begin{figure}
		\includegraphics[width=0.98\linewidth]{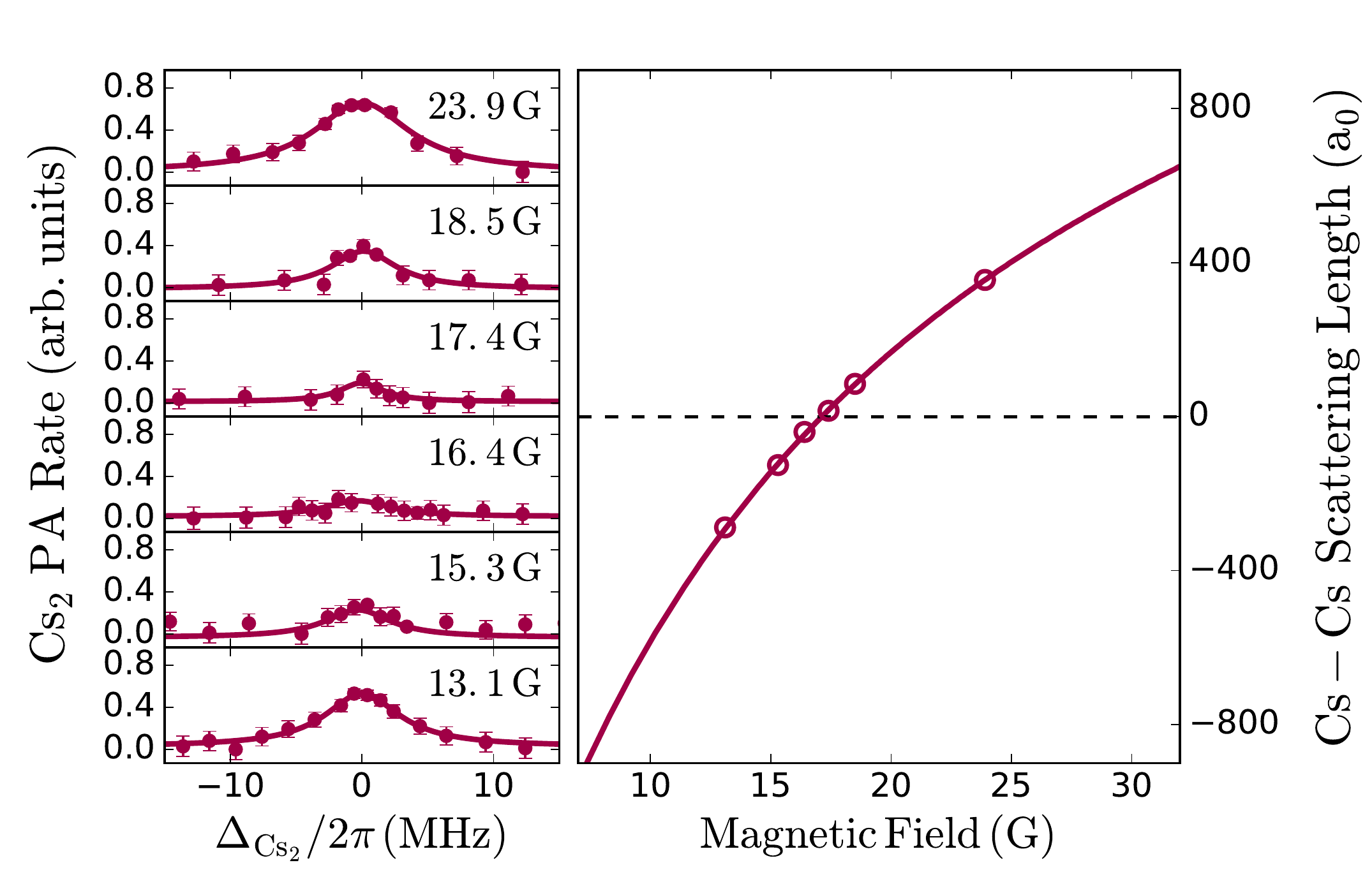}
	\caption{Modification of Cs$_{2}$ photoassociation rate using a Feshbach
resonance. The left panel shows Cs$_{2}$ photoassociation rates as a function
of detuning from the $0^{+}_{u}$ $v=136$ line for varying magnetic field
strengths. The right panel shows the Cs scattering length as a function of
magnetic field \cite{Berninger2013} (for clarity, narrow Feshbach resonances at
$14.4\,$G, $15.1\,$G and $19.9\,$G are not shown). The red circles show the
scattering lengths at magnetic fields corresponding to the measurements on the
left.}
	\label{fig:Cs2}
\end{figure}

Due to the large difference in polarizability at the wavelength of our ODT and
the collision properties of Cs and Yb, we can currently prepare only a mixture with
a large number imbalance in favor of Yb \cite{Guttridge2017}. Therefore,
Cs$^*$Yb PA resonances are detected by loss of Cs atoms from the ODT.
Unfortunately, the Cs atoms are also affected by off-resonant scattering of the
PA light, leading to non-resonant loss and optical pumping into the upper
hyperfine manifold ($F=4$). To improve the signal-to-noise ratio, we use a
pulse of imaging light on the Cs \mbox{$6 S_{1/2}, F=4 \rightarrow 6
P_{3/2}, F'=5$} transition to remove
any atoms off-resonantly pumped into the upper hyperfine level prior to
detection of the population in $\ket{F=3,m_{F}=+3}$.

A larger issue is the existence of the many Cs$_{2}$ PA resonances below the
$D_{1}$ transition \cite{Pichler2004,Pruvost2010,Ma2014,Liu2015,Li2017}, making
identification of CsYb lines challenging. However, due to the tunability of the
scattering length of Cs we can tune the magnetic field to suppress the Cs$_{2}$
PA rate, as shown in Fig.\ \ref{fig:Cs2}. This effect is well understood in the
context of Feshbach-Optimized Photoassociation (FOPA)
\cite{Tolra2003,Pellegrini2008,Junker2008} and is due to the modification of
the scattering wavefunction in the vicinity of a Feshbach resonance which, in
turn, modifies the Franck-Condon overlap with a specific excited vibrational
level. The effect is typically used to enhance the PA rate of a transition. Here,
however, we use the effect to suppress the Cs$_{2}$ PA rate by operating at a
magnetic field of $16.4(2) \,$G when searching for CsYb PA lines. This is not
expected to modify the CsYb PA rate as the predicted Feshbach resonances in
this system are very sparse and narrow \cite{Brue2013}. Note that this magnetic
field properly suppresses Cs$_{2}$ resonances over most of the range of
detunings explored here, but, due to the oscillatory nature of the ground-state
wavefunction, for larger detunings it can enhance the PA rate \cite{Tolra2003}.

We typically measure the CsYb PA lines by illuminating the trapped atomic
mixture with a pulse of PA light for $300 \,$ms at an intensity of $I = 0.1 -
10 \,$W/cm$^{2}$ (depending on the strength of the transition). The ODT light
is then turned off and the number of atoms is measured using resonant
absorption imaging. Short scans (comparable to the cavity FSR) are performed by
tuning the modulation frequency of the fibre EOM, measuring the Cs number with
each frequency step. We stitch together longer scans by locking the PA laser
frequency to sequential modes of the cavity.

\section{Experimental Results}

\subsection{$^{133}$Cs$^{174}$Yb Photoassociation}

\begin{figure}
		\includegraphics[width=0.95\linewidth]{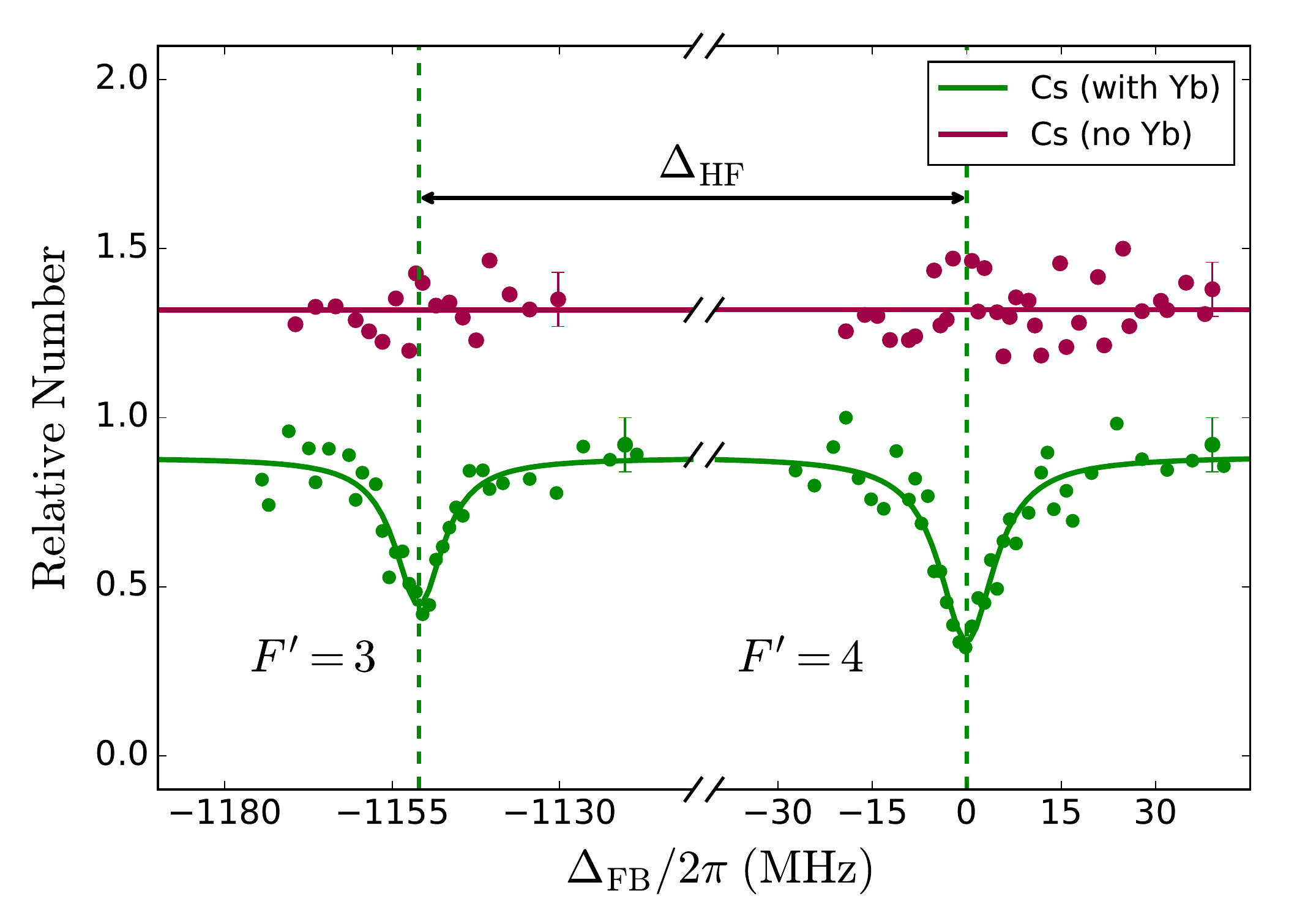}
\caption{Observation of the photoassociation resonance for $n'=-11$ of
$^{133}$Cs*$^{174}$Yb. Relative number of Cs atoms remaining after a 300 ms
pulse of PA light versus detuning from the $F'=4$ line
($\Delta_{\mathrm{FB}}$). The green (red) trace shows the photoassociation
spectra of Cs with (without) the presence of Yb in the dipole trap. The red (Cs
only) trace has been offset for clarity. The statistical error in the atom
number is shown by the error bars on the right hand side. The dashed green line
shows the centres of the CsYb PA resonance for each hyperfine
component.}
	\label{fig:CsYbPA}
\end{figure}

A typical CsYb PA spectrum is shown in Fig.\ \ref{fig:CsYbPA} as a function of
the detuning $\Delta_{\mathrm{FB}}$ from the free-bound transition. The figure
displays the $n'=-11$ line, where we label the lines by numbering the
vibrational levels of the 2(1/2) state below its threshold, starting from
$n'=-1$. Explicitly, $n' = v - v_{\mathrm{max}} -1$, where $v$ is the
vibrational quantum number and $v_{\mathrm{max}}$ is the vibrational quantum
number of the least-bound state. As the levels we observe are all close to
threshold, $n'$ is relatively easy to determine, but we cannot label the states
by $v$ as $v_{\mathrm{max}}$ is initially unknown.

When the frequency of the PA laser is tuned into resonance with a CsYb line we
observe a loss of Cs atoms due to the formation of Cs$^*$Yb molecules. We
verify that the detected features are CsYb resonances (and not Cs$_2$
resonances) by repeating the scan in the absence of Yb. To keep the density and
temperature of the Cs atoms comparable to the measurement taken with Yb, we
simply remove Yb from the ODT with a pulse of light resonant with the
$^{1}S_{0} \rightarrow\, ^{1}P_{1}$ transition immediately before the sample is
illuminated by the PA light. The disappearance of the feature in the absence of
Yb (red trace in Fig.\ \ref{fig:CsYbPA}) confirms the existence of a CsYb PA
resonance.

For all vibrational levels we observe a second PA feature which is red-detuned
by approximately the hyperfine splitting of the Cs $6 P_{1/2}$ level. For the
weakly bound vibrational states investigated here, the Cs$^*$Yb molecules
inherit the properties of the two free atoms; as such we identify the two lines
by the quantum numbers $F'=4$ and $F'=3$ corresponding to the hyperfine
structure in the excited state of Cs. The rovibrational levels are best
described by the classic form of Hund's case (e) introduced by Mulliken
\cite{Mulliken:1930}, in which the total atomic angular momentum ($F'$ here)
couples to the rotational angular momentum $R'$ to form a resultant ${\cal
F}'$. This uncommon coupling case was first observed for HeKr$^+$
\cite{Carrington:hekr:1996} and has also been found in RbYb
\cite{Nemitz2009,Bruni2016}. In our case, all rovibrational levels observed
have $R' = 0$ because of the low temperature of the initial atomic mixture.

\begin{table}
	\begin{center}
	\begin{ruledtabular}
	\begin{tabular}{ >{\centering\arraybackslash}m{1.5cm} >{\centering\arraybackslash}m{2.5cm} >{\centering\arraybackslash}m{2cm} >{\centering\arraybackslash}m{2.5cm}}
	
 \multicolumn{1}{m{1.5cm}}{\centering $n'$}
&\multicolumn{1}{m{2.5cm}}{\centering $\Delta_{\mathrm{PA}}/2\pi$  \\ (GHz)}
&\multicolumn{1}{m{2cm}}{\centering Normalized  \\ strength}
&\multicolumn{1}{m{2.5cm}}{\centering $\Delta_{\mathrm{HF}}/2\pi$  \\ (MHz)}\\
	\hline
	Cs & 0 & N/A & 1168(2)\\
	\hline
	-7 & -17.244(3) & 1.0(2) & 1162(1) \\
	-8 & -26.473(3) & 0.4(3) & 1157(3) \\
	-9 & -38.567(3)& 0.40(5)& 1154(1) \\
	-10 & -53.932(3)& 0.17(1)& 1151(1) \\
	-11 & -72.973(3)&0.19(1)& 1147(1) \\
	-12	& -96.091(3)&	0.091(8)& 1142(2) \\
	-13	& -123.678(3)& 0.10(2)& 1139(1) \\
	-14	& -156.117(3)& 0.045(4)& 1131(1)\\
	-15	& -193.772(3) & 0.06(2)& 1127(1)\\
	-16	& -236.991(3)& 0.013(2)& 1120(1)\\
	-17 & -286.098(4)&0.05(1)& 1115(2)\\
	-18	& \multicolumn{3}{c}{--------- not observed ---------}\\
	-19	& -402.867(8)& 0.0063(4)& 1071(8)\\
	-20 & -472.384(6)& 0.0033(6)& 1084(6)\\
	\end{tabular}
	\end{ruledtabular}
	\end{center}
	\caption{Measured binding energies of vibrational levels in the 2(1/2)
electronically excited state of $^{133}$Cs$^{174}$Yb. Binding energies are
given for the $F'=4$ level and are measured relative to the Cs $D_{1}$ atomic
transition $\ket{6 S_{1/2}, F=3 \rightarrow 6 P_{1/2}, F'=4}$. The
uncertainties quoted are $1 \sigma$ uncertainties \cite{Hughes2010}. The
observed strengths of the lines are normalized to that of the strongest PA
line, $n' = -7$. The hyperfine splittings are the measured separations of the
$F'=4$ and $F'=3$ components. The measured atomic value is in agreement with
the literature value of the hyperfine splitting of the $m_F=+3$ levels in a
$2.2 \,$G magnetic field, 1169.272(81) \cite{Udem1999}. }\label{table:Binding
energies}
\end{table}

Table \ref{table:Binding energies} lists the binding energies of all observed
vibrational levels relative to the Cs \mbox{$6 S_{1/2}, F=3 \rightarrow 6
P_{1/2}, F'=4$} atomic transition. These measurements were performed at a
magnetic field of $2.2(2) \,$G to reduce uncertainty caused by the Zeeman shift
of the molecular state when measuring the hyperfine splitting. The binding
energies are obtained using the difference in EOM modulation frequencies and
number of cavity FSRs between the PA transition and the atomic transition, as
outlined earlier. The uncertainty due to the stabilisation of the cavity length
is the dominant source of uncertainty for the majority of the measured binding
energies. The exception is the $n' = -19$ line, where the observed FWHM
linewidth of $130(10) \,$MHz leads to a larger uncertainty in determining the
line centre. All the other features have linewidths approximately equal to the
linewidth of the Cs $D_{1}$ transition, as shown in Fig.\ \ref{fig:CsYbPA}.

The strength of the transition is determined by observing the loss of Cs atoms
as a function of intensity of PA light. We observe an exponential decay of the
Cs atom number as a function of intensity. The decay constant extracted from
the exponential fit is normalized to that of the $n' = -7$ level and given in
Table \ref{table:Binding energies}.

\begin{figure}
		\includegraphics[width=0.95\linewidth]{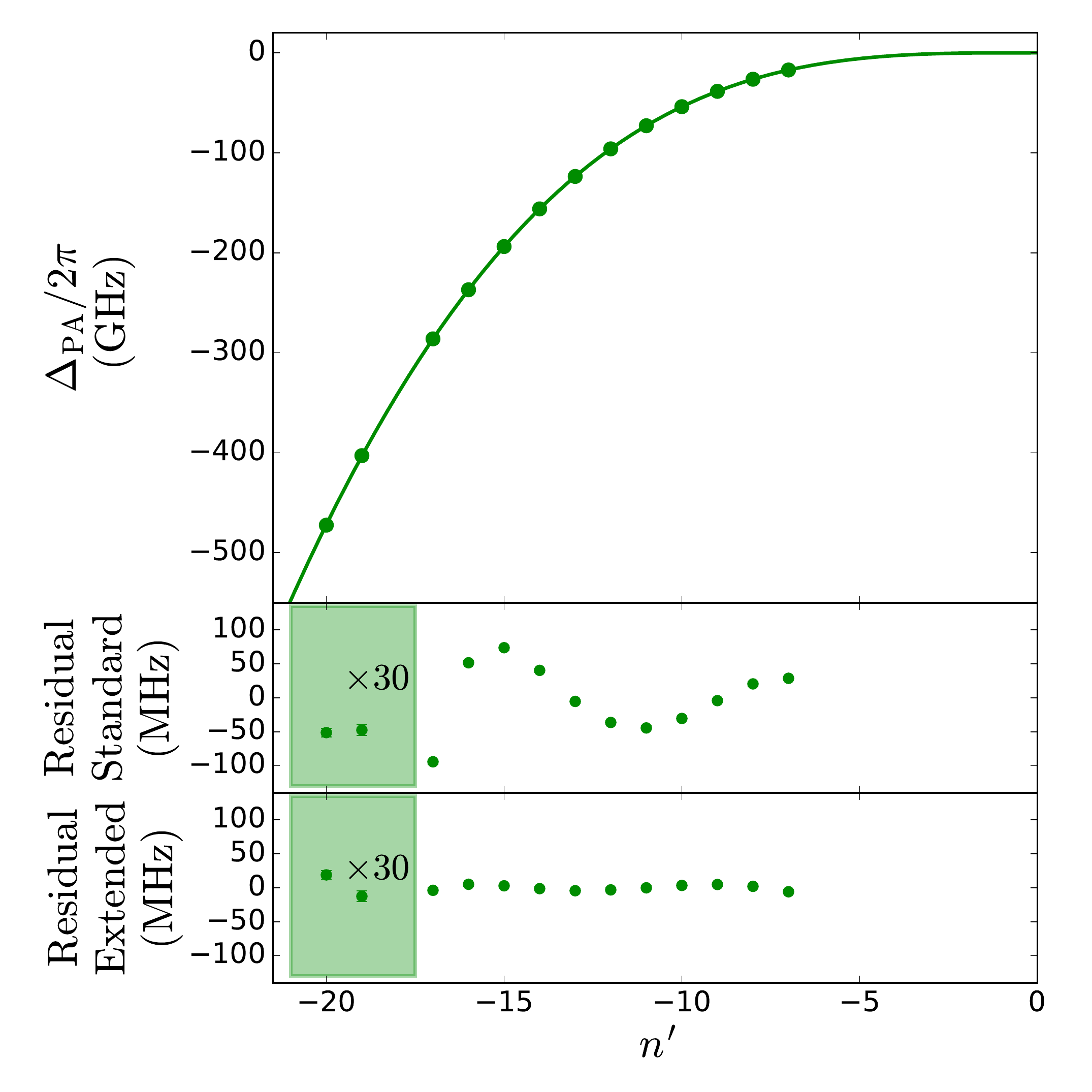}
	\caption{Binding energies of vibrational levels of $^{133}$Cs$^{174}$Yb
on the 2(1/2) excited-state potential by one-photon photoassociation
spectroscopy. Upper: The detuning of observed $F'=4$ PA resonances are plotted
against the vibrational quantum number counted from dissociation, $n'$. The
dissociation energy corresponds to the $\ket{6 S_{1/2}, F=3 \rightarrow 6
P_{1/2}, F'=4}$ atomic transition. The solid green line shows a fit to the data
using the extended Le\,Roy-Bernstein equation (see text). The lower two panels
compare the residuals for the fits using the standard and extended Le\,Roy
Bernstein equations. There is clear discrepancy for $n'=-19$ and $n'=-20$ whose
residuals are 30 times larger than those plotted on the figure. Most of the
error bars are much smaller than the data points.}
	\label{fig:LeRoy}
\end{figure}

Figure \ref{fig:LeRoy} shows the measured binding energies of all one-photon PA
transitions found for $^{133}$Cs$^{174}$Yb. Vibrational levels are observed
with binding energies of order 20 to 500 GHz detuned from the Cs $D_{1}$ line.
These measured binding energies are relatively small compared to the depth of
the potential ($ \approx 200$ THz), so the positions of the vibrational levels
are determined by the long-range potential. The potential curve for a pair of
atoms can be described at sufficiently large internuclear distance $R$ by an
inverse-power series
\begin{equation}
V(R) = D - \frac{C_{n}}{R^{n}} - \frac{C_{m}}{R^{m}} - ...,
\end{equation}
where $V(R)$ is the potential as a function of internuclear distance, $D$ is
the threshold energy, and $C_n$ and $C_m$ are long-range coefficients. At long
range, the CsYb 2(1/2) potential is dominated by the van der Waals $n=6$ term.
The long-range coefficients may be extracted from PA spectra using
near-dissociation expansion formulas. The simplest and most widely used of
these expansions is the Le\,Roy-Bernstein (LRB) formula \cite{LeRoy1970} which
links the energy $E_v$ of the vibrational state $v$ to the asymptotic form $(D
- C_{n}/R^{n})$ of the potential
\begin{equation}
E_v \simeq D -\left(\frac{v_{\rm{D}}-v}{B_n} \right)^{2n/(n-2)},
\end{equation}
where $v_{\mathrm{D}}$ is the non-integer vibrational quantum number at
dissociation and $B_n$ is a constant that depends on the reduced mass and the
leading long-range power $n$. In practice, it is more convenient to express
$v_{\rm{D}}-v$ in terms of $n'$ and $v_{\rm{frac}}$, the fractional part of
$v_{\rm{D}}$; for a single isotope, $v_{\mathrm{max}}$ does not affect the predicted level positions.

In searching for PA lines, we modelled our data using the LRB equation (for
$n=6$) and used the fitted parameters to predict more deeply bound levels. This
technique yielded accurate predictions for levels up to $n' = -17$, with the
measured binding energies typically lying within a few hundred MHz of the
predicted values. For the more deeply bound levels $n' = -19$ and $n' = -20$,
the measured line frequencies were far from the extrapolated values and the $n'
= -18$ level was not observed at all. The non-observation of the $n' = -18$
level may be due to a small Franck-Condon factor or that the level is located
outside the range searched ($\pm 2 \,$GHz from the prediction) or coincided
with a Cs$_2$ transition. We did not search further due to the $\sim 30$\,s
load-detection cycle associated with conducting the measurements.

The middle panel of Fig.\ \ref{fig:LeRoy} shows the residuals from the fit of
our PA measurements to the LRB equation. The $n'=-19$ and $n'=-20$ levels are
outliers and so not included in any of our fits. It is clear from the residuals
that the standard LRB equation does not fully describe our measured PA spectra.
The structure of the residuals suggests that a model including higher-order
terms would give a better fit to the results. Indeed, the more strongly bound
levels with binding energies around 300 GHz are deep enough to be sensitive to
the non-asymptotic, short-range character of the potential for our measurement
precision.

To model the PA spectra better, we also fit them using an extended version of
the LRB equation, specifically Eq. (39) in Ref \cite{Comparat2004}. The
extended version allows the inclusion of one higher-order dispersion
coefficient (we use $m=8$) and a mass-dependent parameter $\gamma$ which
accounts for the non-asymptotic, short-range part of the potential. The bottom
panel of Fig.\ \ref{fig:LeRoy} shows the residuals of the fit to the extended
LRB equation. The inclusion of the extra terms significantly improves the fit
to the results. The reduced chi-squared of the extended fit is $\chi^{2}_{\nu}
= 2.3$, much better fit than the standard LRB equation which gives
$\chi^{2}_{\nu} = 275$. The best-fit parameters for the extended fit are $C_{6}
= 10.1(1) \times 10^{3} \, E_{\rm h} a_0^6$, $C_{8} = 5.0(2)\times 10^{6}
\,E_{\rm h} a_0^8$, $v_{\mathrm{frac}} = 0.696(6)$ and $\gamma^{-1}=h\times
3.4(1) \times 10^{2}$~GHz.

When fitting to either model, the residuals for $n' = -19$ and $-20$ are over
30 times larger than that of the other levels. These levels may be perturbed by
mixing with vibrational levels in a different electronic state
\cite{Borkowski2014}. The shift could also be caused by the broadband dipole
trapping light coupling to a higher electronic state. The $n' = -19$ line is
extremely broad in comparison to other observed lines; it has a FWHM of
$130(10) \, \mathrm{MHz}$, over eight times the linewidth of $n' = -16$ (FWHM $
=15(2) \, \mathrm{MHz}$) at the same light intensity. We have not been able to
observe any levels beyond $n' = -20$, although we have searched a moderate $\pm
1 \,$GHz range around the predicted positions. As can be seen from the
residuals for the deepest observed states in Fig.\ \ref{fig:LeRoy}, the
disagreement with the LRB fit results in an increasingly large search space,
which is very time consuming to explore.

\begin{figure}
		\includegraphics[width=0.95\linewidth]{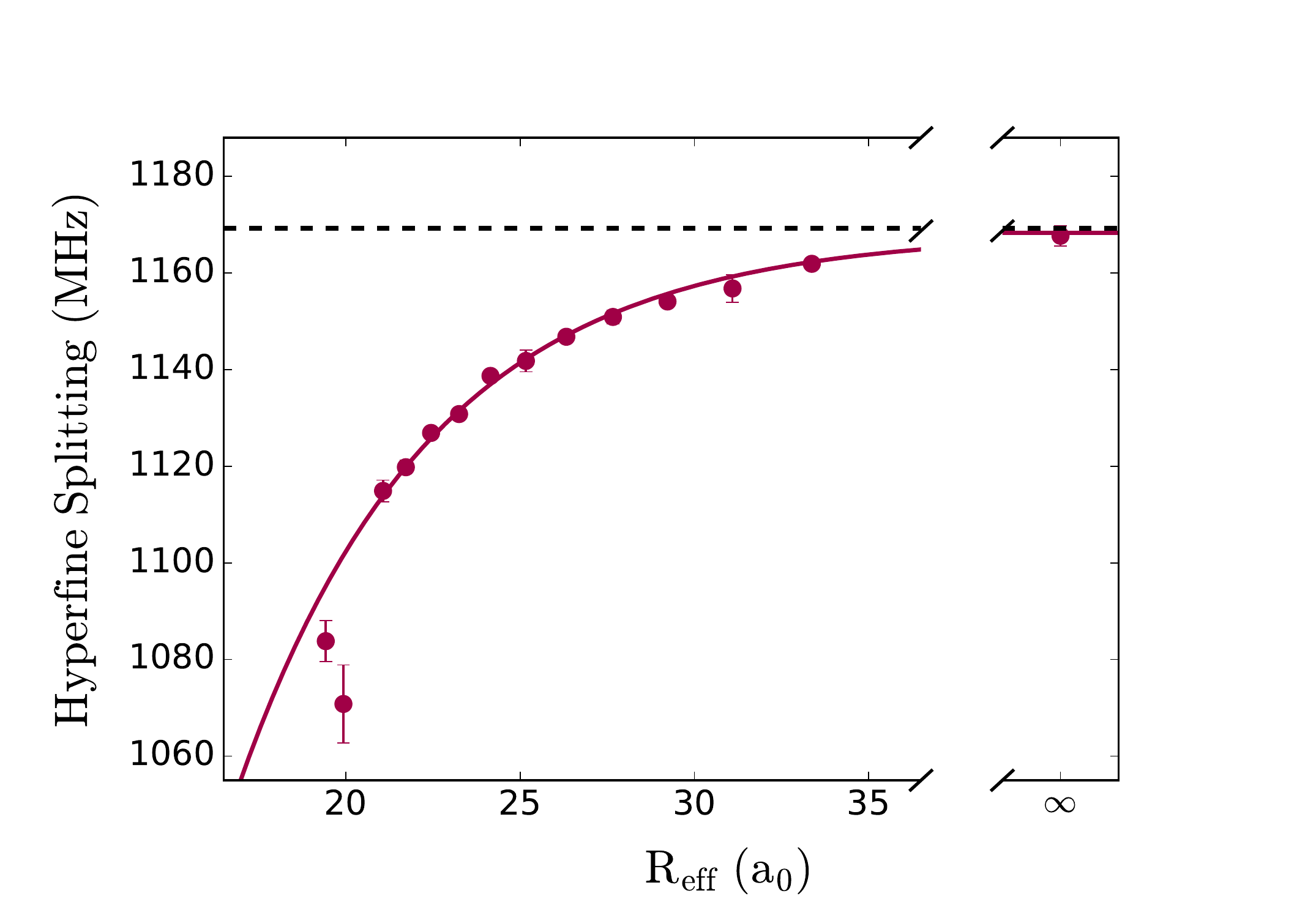}
	\caption{Measured hyperfine splitting as a function of the effective internuclear
distance, $R_{\rm{eff}}$ (see text for details). The horizontal dashed line shows
the Cs atomic hyperfine splitting of the $m_F=+3$ levels in the $6 P_{1/2}$ state at a magnetic
field of $2.2(2) \,$G  \cite{Udem1999,Steck2010b}. The solid line is an exponential fit to guide the eye.}
	\label{fig:Hyperfine}
\end{figure}

As discussed earlier, the CsYb spectra display hyperfine structure associated
with the Cs atom (see Fig.\ \ref{fig:CsYbPA}). We present the measured
hyperfine splitting for all the observed levels in Table~\ref{table:Binding
energies} and we illustrate the dependence of the hyperfine coupling on
internuclear distance in Fig.\ \ref{fig:Hyperfine}. We approximate the
effective internuclear distance $R_{\rm eff}$ for each transition as the Condon
point, where the transition energy is equal to the spacing between the two
curves. The points show the measured hyperfine splitting of the $F'=4$ and $F'=3$ levels of each vibrational level. We find that as the binding energy increases and the internuclear separation reduces, the strength of the Cs hyperfine coupling decreases. This is due to the perturbation of the electronic wave function of the Cs atom by the presence of the closed-shell Yb atom \cite{Zuchowski2010}. A similar effect in the ground state has been observed to produce Feshbach resonances in RbSr \cite{Barbe2017}.
The deepest bound level $n' = -20$ exhibits a hyperfine splitting of
$\Delta_{\mathrm{HF}} = 1084(6) \,$MHz, a reduction of almost $100 \,$MHz from
the atomic value. The hyperfine splitting of $n' = -19$ is smaller than this
value but this may be due to mixing of vibronic states in other electronic
states causing a modification of the coupling.

\subsection{Extension to other CsYb Isotopologs}

Ytterbium has numerous stable isotopes, both bosonic and fermionic, that can be
trapped and cooled to ultracold temperatures
\cite{Takasu2003,Fukuhara2007a,Fukuhara2007,Takasu2009,Taie2010,Sugawa2011}.
Within the Born-Oppenheimer approximation, the interaction potential is
mass-independent but the positions of vibrational levels depend on the reduced
mass.

In WKB quantization, the non-integer quantum number at dissociation,
$v_{\mathrm{D}} = v_{\mathrm{max}} + v_{\mathrm{frac}}$, is given by
$v_{\mathrm{D}} = \Phi/\pi - 1/2$, where $\Phi$ is the phase integral
\begin{equation}
\Phi = \int^{\infty}_{R_{\rm in }} [(2 \mu/\hbar^2) (D-V(R))]^{1/2} \, dR.
\end{equation}
Here $R_{\rm{in}}$ is the location of the inner classical turning point, $\mu$
is the reduced mass and $V(R)$ is the interaction potential. The dependence on
$\mu$ allows us to determine the number of bound states $N=v_{\rm max}+1$ by
comparing binding energies for different isotopologs.

\begin{table}
	\begin{ruledtabular}
	    \begin{tabular}{ccccccc}
    Yb & \multirow{2}{*}{$n'$} &  $\Delta_{\mathrm{PA}}/2\pi$   & \multicolumn{2}{c}{Residual (MHz)} \\
    Isotope &             &     (MHz)   & $N=154$ & $N=155$  \\
    \hline

	
	173 & -9 & -36.117(3)& -10&6\\
	173 & -10 & -50.877(3) & -10&11\\
	173 & -11 & -69.246(3) & -15 &10\\
	173 & -12	& -91.633(3) & -19 &12\\
	173 & -13	& -118.427(3) & -23&13 \\
	173 & -14	& -150.014(3) & -29&14\\
	173 & -15	& -186.762(3) & -38&11\\ \hline
	
	172 & -8	& -22.740(5) & -14&10\\
	172 & -11	& -65.614(4) & -21&28\\
	172 & -13	& -113.258(4) & -45&26\\ \hline
	
	170 & -12	& -103.338(3) & 15&150\\
	170 & -14 & -166.489(3) & 55 &241\\

	\end{tabular}
	\end{ruledtabular}
	\caption{Measured binding energies of vibrational levels in the 2(1/2)
molecular potential for different isotopologs of $^{133}$CsYb. The binding
energies quoted are for the $F'=4$ level and are measured relative to the Cs
$D_{1}$ atomic transition $\ket{6 S_{1/2}, F=3 \rightarrow 6 P_{1/2}, F'=4}$.
The residuals presented are from the extended LRB model with $N=154$ or $N=155$.}	
	\label{table:Isotopologs}
\end{table}

The measured binding energies of $^{133}$Cs$^{173}$Yb, $^{133}$Cs$^{172}$Yb and
$^{133}$Cs$^{170}$Yb are tabulated in Table \ref{table:Isotopologs}. The
routines used to obtain PA spectra for these isotopologs are similar to that
presented for $^{133}$Cs$^{174}$Yb, with the only significant difference in the
preparation of the ultracold Yb sample. Slight changes are required to the MOT,
ODT loading and evaporative cooling routines to address the different
requirements of each Yb isotope due to variations in abundance, intraspecies
scattering length and hyperfine structure (for fermionic $^{173}$Yb). The
$^{133}$Cs$^{173}$Yb and $^{133}$Cs$^{170}$Yb measurements take place in
identical trapping conditions to $^{133}$Cs$^{174}$Yb. The initial mixture
contains \mbox{$3 \times 10^{5}$ $^{173}$Yb} or  \mbox{$4 \times 10^{5}$
$^{170}$Yb} atoms at $T_{\mathrm{Yb}} = 1 \, \mu$K and \mbox{5 $\times 10^{4}$}
Cs atoms at $T_{\mathrm{Cs}} = 6 \, \mu$K. The large negative scattering length
of $^{172}$Yb ($a_{172-172} =-600 \ a_{0}$) \cite{Kitagawa2008} complicates the
evaporative cooling of Yb; we therefore halt the evaporation around
$T_{\mathrm{Yb}} = 4\  \mu K$ to prevent a substantial loss of Yb atoms due to
3-body inelastic collisions. PA for $^{133}$Cs$^{172}$Yb is performed on a
mixture of $5 \times 10^{5}$ $^{172}$Yb atoms at $T_{\mathrm{Yb}} = 4 \, \mu$K
and \mbox{7 $\times 10^{4}$} Cs atoms at $T_{\mathrm{Cs}} = 12 \, \mu$K. In
this new trapping arrangement the Yb (Cs) trap frequencies are 380 (1100) Hz
radially and 80 (240) Hz axially. The light shift due to this tighter
trapping arrangement has been accounted for in the binding energies presented
in Table \ref{table:Isotopologs} and leads to a larger uncertainty on the
$^{172}$Yb measurements.

To determine $N$ from the measured binding energies of the four isotopologs we
use a mass-scaled version of the extended LRB model. The values of $C_{6}$ and
$C_{8}$ are the same for all isotopologs. However, $v_{\rm D}$ is proportional
to $\sqrt{\mu}$, and so $v_{\rm frac}$ varies between isotopologs. $\gamma$ is
also proportional to $\sqrt{\mu}$ \cite{Comparat2004}, but this variation is
much less important than that for $v_{\rm frac}$. For a chosen value of $N$, we
can use the parameters fitted to Cs$^{174}$Yb to predict binding energies for
the other isotopologs and calculate $\chi^2_\nu$. It is possible to refit the
parameters with multiple isotopologs, but this makes little quantitative
difference and produces the same qualitative conclusions.

The binding energies for Cs$^{172}$Yb and Cs$^{173}$Yb are well predicted by
the parameters obtained for Cs$^{174}$Yb with $N=155$, giving $\chi^2_\nu=12$.
This compares with $\chi^2_\nu=40$ and 158 for $N=154$ and 156 respectively.
However, including Cs$^{170}$Yb gives $\chi^2_\nu=36$ for $N=154$ and 322 for
$N=155$. It thus appears that the results for the different isotopologs are
inconsistent with a single-potential model; the deviations are outside the
experimental errors and clearly non-statistical.

It is possible that the lines for one or more isotopes are affected by an
isotope-dependent perturbation, most likely due to a level of the 3(1/2)
electronic state that dissociates to the $6\,^2P_{3/2}$ state of Cs. Such a
perturbation is not encapsulated in our model and characterizing it would
require extensive further work. Nevertheless, we can conclude that the number
of bound states supported by the 2(1/2) potential is either 154 or 155. This is
within $10\%$ of the $145$ bound states predicted for this potential by
Meniailava and Shundalau \cite{Meniailava2017}.

\section{Conclusion}
We have produced ultracold Cs$^*$Yb molecules using photoassociation on an
atomic mixture trapped in an optical dipole trap. We have measured the binding
energies of 13 vibrational levels of the electronically excited 2(1/2) state of
$^{133}\rm{Cs}$$^{174}$Yb and fitted dispersion coefficients for its
interaction potential at long-range near the Cs $6\,^2P_{1/2}$ asymptote. The low
temperatures and well-defined internal states of the atoms in the optical
dipole trap allow us to measure the hyperfine splitting of the molecules
associated with the Cs $6\,^2P_{1/2}$ state. For more deeply bound Cs$^*$Yb
molecules we observe a decrease in the hyperfine splitting compared to the bare
Cs atom. In addition, we measure the binding energies of a number of
vibrational levels of $^{133}$Cs$^{173}$Yb, $^{133}$Cs$^{172}$Yb and
$^{133}$Cs$^{170}$Yb. By applying mass scaling, we determine the number of
bound states supported by the 2(1/2) potential of $^{133}$Cs$^{174}$Yb, which
correlates at short range with the $1\,^2\Pi_{1/2}$ potential, to be 154 or
155. $^{133}$Cs$^{173}$Yb and $^{133}$Cs$^{172}$Yb also have this number of bound states, but $^{133}$Cs$^{170}$Yb has one fewer bound states.

The improved understanding of the electronically excited state will be pivotal
in the creation of ground-state CsYb molecules. The measurements presented here
are the starting point for two-photon photoassociation to near-threshold levels
of the $X ^{2}\Sigma^{+}_{1/2}$ ground-state potential and for all-optical
approaches such as STIRAP to produce molecules in the absolute ground state
\cite{Stellmer2012,Ciamei2017}. Two-photon PA will also allow precise
determination of the interspecies scattering lengths and the prediction of
Feshbach resonances suitable for magnetoassociation. Ground-state CsYb
molecules may find future applications in the fields of ultracold chemistry,
precision measurement and quantum simulation.

\begin{acknowledgments}
We acknowledge support from the UK Engineering and Physical Sciences Research Council (grant number EP/P01058X/1). JJM acknowledges an International Engagement Travel Grant from Durham University. The data presented in this paper are available from \url{http://dx.doi.org/10.15128/r16d56zw600}.
\end{acknowledgments}

\bibliography{PhotoassociationReferences}

\end{document}